\documentclass[10pt,letterpaper]{article}
\usepackage[top=0.85in,left=1.6in,footskip=0.75in]{geometry}

\usepackage{amsmath,amssymb}

\usepackage{changepage,physics}

\usepackage[utf8x]{inputenc}

\usepackage{textcomp,marvosym}

\usepackage{cite}

\usepackage{nameref,hyperref,soul}

\usepackage[right]{lineno}

\usepackage{microtype}
\DisableLigatures[f]{encoding = *, family = * }

\usepackage[table]{xcolor}

\usepackage{array}

\usepackage{caption,tabularx,booktabs}

\newcolumntype{+}{!{\vrule width 2pt}}

\newlength\savedwidth

\usepackage[aboveskip=1pt,labelfont=bf,font=small,labelsep=period,singlelinecheck=off]{caption}

\makeatletter
\renewcommand{\@biblabel}[1]{\quad#1.}
\makeatother

\usepackage{lastpage,fancyhdr,graphicx}
\usepackage{epstopdf}
\begin{document}
\vspace*{0.2in}

\begin{flushleft}
{\Large
\textbf\newline{\textbf{Economical representation of spatial networks}}
}
\newline
\\
Fabrizio De Vico Fallani\textsuperscript{1*},
Thibault Rolland\textsuperscript{1}
\\
\bigskip
\textbf{1}  Sorbonne University, Paris Brain Institute (ICM), CNRS, Inria,
Inserm, AP-HP, Pitie-Salpetriere Hospital, Paris, France
\bigskip

* corresponing author: fabrizio.de-vico-fallani@inria.fr

\end{flushleft}
\bigskip

\section*{Abstract}

\begin{small}
Network visualization is essential for many scientific, societal, technological and artistic domains. The primary goal is to highlight patterns out of nodes interconnected by edges that are easy to understand, facilitate communication and support decision-making.
This is typically achieved by rearranging the nodes to minimize the edge crossings responsible of unintelligible and often unaesthetic trends. But when the nodes cannot be moved, as in spatial and physical networks, this procedure is not viable.
Here, we overcome this situation by turning the edge crossing problem into a graph filtering optimization. 
We demonstrate 
that the presence of longer connections prompt the optimal solution to yield sparser networks, thereby limiting the number of intersections and getting more readable layouts.
This theoretical result matches human behavior and provides an ecologically-inspired criterion to visualize and model real-world interconnected systems.

\end{small}


\newpage

\section*{Introduction}

In the era of big data, representing complex information in a visually concise and effective manner is crucial to ease communication and decision-making. 
In an increasingly interconnected world, network visualization plays a fundamental role in identifying intelligible trends within complex diagrams made up of nodes linked by lines.
The number of potential applications is huge, spanning various fields from mathematics and biology to technology and art \cite{lima_visual_2011,von_landesberger_visual_2011,filipov_are_2023}.

While there are no strict criteria for improving the readability of a network, it is generally agreed that the corresponding drawing should have minimal edge crossing, with nodes evenly distributed in the space, connected nodes close to each other, and symmetry that may exist in the graph preserved \cite{purchase_validating_1996,purchase_metrics_2002,chen_evaluating_2021}.
To this end, many algorithms have been developed based on different criteria such as the spring-electrical models, the stress and strain models, a well as high-dimensional embedding and Hall’s algorithms \cite{herman_graph_2000}. 
The main working strength of all these methods is the possibility to freely rearrange the node positions so as to optimize some quality function associated with the human perception.

Yet, in many real-world systems such as spatial and physical networks, the precise positioning of the nodes cannot be altered without losing information about the system's intrinsic geometry.
If rearranging the nodes is not possible, the alternative is to focus on the links, for example by bending or stretching their shape to minimize intersections \cite{tamassia_handbook_2016}. However, these networks might be difficult to comprehend because of the inevitable link tortuosity. 
In addition, while in three dimensions this approach can result in wiring patterns devoid of any crossing \cite{dehmamy_structural_2018}, artificial intersections persist due to the presence of multiple overlapping plans.

In general, edge crossings exponentially increase with the connection density. Notably, the longer are the connections, the higher is the likelihood of having multiple intersections. 
Yet, the network connections constitute basic informative units. The greater their number, the more detailed is our knowledge of the system organization.
Hence, the problem of network representation can be remapped into a graph filtering optimization balancing the benefit of including as many connections as possible and the incurred cost due to their length, an indirect proxy of edge crossings.

By solving the associated analytical formulation, we reveal a nontrivial relationship between the optimal connection density and the spatial distribution of the edges within the network.
We confirm this theoretical behavior using data from human responses collected from an online interactive experiment involving $n=10687$ participants. Based on the gathered answers, we derive an unbiased criterion to filter networks and get readable representations of otherwise too dense real interconnected systems.

Drawing from these principles, we eventually introduce a benefit-cost network model that produces a continuous spectrum of realistic configurations and evaluate its ability to reproduce the spatial and topological properties of the \textit{C.elegans}' neuronal network.
 

%

\section*{Results}
\subsection*{Graph filtering}
A generic network can be mathematically described by a graph with  $N$ nodes and $L$ links, or edges. Here, we considered undirected weighted graphs where each node is further equipped with a position in a Euclidean \textit{s}-dimensional space.
A basic characterization of such geometric graph is given by two quantities, i.e. the connection density $\rho=\frac{2L}{N(N-1)}$ and the spatial
density $\delta$, defined as the cumulative length of the $L$ edges divided by the maximum when the graph is fully connected (\textbf{Text S1}).

To measure the balance between the above quantities, we considered the general functional $J=f(\rho) g(\delta)$ where $f$ and $g$ are respectively increasing and decreasing functions. Because both $\rho$ and $\delta$ are in the $[0, 1]$ interval, a very simple form reads 

\begin{equation}
J= \rho^\alpha(1-\delta)^\beta
\end{equation}

where $\alpha \ge 0$ and $\beta \ge 0$ are parameters tuning the importance of the connection and spatial density, respectively.
Using the binomial approximation, Eq. 1 can be rewritten in terms of benefit-cost $J\simeq \rho^\alpha-\rho^\alpha \beta \delta$ indicating that the cost per length unit
$c=\rho^\alpha \beta$ is not constant, but grows with the number of existing connections in the network. This behavior naturally reflects the fact that in denser graphs, the edges have a higher likelihood to produce several crossings.

Our goal was to find the optimal number of links, or equivalently the optimal connection density $\rho$, that maximizes $J$. By construction, the cumulative length, here measured by the internode Euclidean distance $d$, increases with the number of links. It is therefore convenient to find a formal relation between $\delta$ and $\rho$. 
Let us assume that the internode distances fall in the unitary interval, the link weights are positive and that both are unique\footnote{Link weights typically measure some structural or functional property of the system. If this information is missing, the actual internode distances can be used here as a proxy for the weights.}.
When the weights are randomly distributed between the nodes, they do not correlate with the actual distances. Put differently, selecting the links by their weight corresponds to randomly sample the distances. The probability to choose an edge with a given length is uniform and it is trivial to prove that the expected spatial density $\delta=\rho$ (\textbf{Text S1}).
However, in many real situations the edges' weights and lengths are correlated favoring the emergence of strong \textit{short}-range or \textit{long}-range configurations \cite{}.
Assuming perfect correlation, selecting the links by their weight will correspond to selecting the distances in the same or reverse order.
Hence, the probability to pick an edge with a given length is not uniform but depends on its position in the ranking. Leveraging tools from order statistics, we show that the expected spatial density $\delta \simeq \rho^2$, or $\delta \simeq 2\rho-\rho^2$, depending on whether the strongest edges connect the closest or farther nodes (\textbf{Text S1}).

Considering this space-connection dependency, one can momentarily discard the parameter controlling the edge length (i.e., $\beta=1$) and obtain a simplified one-parameter functional 
$J=\rho^\alpha -\rho^\alpha\delta$.
By substituting the above expressions in the last equation and solving $\dv{J}{\rho}=0$, we obtain the optimal connection density for the three characteristic scenarios

\begin{equation}
  \rho_{long}=\dfrac{\alpha}{\alpha+2},   \hspace{6mm} \rho_{rand}=\dfrac{\alpha}{\alpha+1},\hspace{6mm}  \rho_{short}=\sqrt{\dfrac{\alpha}{\alpha+2}}.
\end{equation}

Eqs. 2 indicate that the optimal number of links solely depends on their spatial distribution. Denser networks emerge when links tend to connect closer nodes, while sparser networks result when links tend to connect farther nodes. The solutions for long-range and short-range configurations establish the theoretical boundaries, with any other arbitrary network falling somewhere in between, i.e., $\rho_{long}\leq \rho_{rand} \leq \rho_{short}$. 
The overall filtering can be modulated by the parameter $\alpha$, giving structural transitions from sparser to denser graphs.
For randomly assigned links, the critical point $\alpha=1$ returns networks with the maximal amount of information in terms of Shannon entropy (i.e. $\rho=0.5$). Notably, when $\alpha<1$ the corresponding optimal densities tend to the values of the long-range case, while for $\alpha>1$ they tend to the short-range one.
For a given connection density, larger $\alpha$ values will preferentially filter long-range connections, while smaller values will rather keep short-distance ones (\textbf{Fig 1}). 


\subsection*{Network visualization}

Graph filtering offers an effective solution to improve the readability of networks whose nodes cannot be arbitrarily rearranged to avoid line criss-cross. 
Long-range connections are particularly problematic as they typically intersect several edges giving unintelligible, often unaesthetic, patterns. 
%
%
Moreover, real networks are typically represented on physical supports with limited space. This means that the larger is the network, the more difficult is to obtain a readable pattern because of the node concentration. 
%
%
To compensate this effect, a natural solution is to let the filtering parameter scale with the typical internode distance $\alpha=\phi N^{-1/s}$, where $\phi$ is a positive constant \cite{moltchanov_distance_2012}. 
%
%
By substituting the latter in Eq. 2, the optimal density in sufficiently big networks becomes $\rho_{long} = \frac{\phi}{2} N^{-1/s}\leq \rho_{rand}  = \phi N^{-1/s} \leq \rho_{short} = \sqrt{\frac{\phi}{2}} N^{-1/2s}$ (\textbf{Text S1}). These expressions preserve the original filtering behavior, 
but now scale with the inverse of the network size thus facilitating the visualization of large systems (\textbf{Fig 1} inset).

How to choose the constant $\phi$ in a possible unbiased way remains unknown. Without any theoretical ground, we turned this fundamental question from a human perception perspective. We realized a simple online experiment involving more than $10000$ trials from different individuals. For each trial, a fixed number of nodes was prompted on pseudo-random 2D grid and the participant was asked to keep adding edges via an interactive slider until the graph become too confusing  (\textbf{Methods}). The edge spatial distribution varied randomly across trials, enabling the added links to connect first either \textit{i)} the farthest nodes, \textit{ii)} the closest ones, or \textit{iii)} in a random fashion (\textbf{Fig 2a}). 
Results show a general preference for very sparse networks regardless of the spatial ordering of the links and a relatively high inter-subject variability \textbf{Fig 2b}). This behavior can be accurately explained by a stochastic Gamma-Poisson process (\textbf{Fig 2b} left inset, \textbf{Text S1}). 
Despite such heterogeneity, the connection density chosen for short-range configurations was statistically higher than that obtained in random and long-range ones (Cohen's $d>0.6$, \textbf{Fig 2b} right inset). 
By fitting Eqs 2 to the actual mean $\rho$ values we eventually derived an unbiased estimate of the filtering constant, i.e., $\phi^*\simeq 1$ (\textbf{Methods}, \textbf{Fig 2b} inset). 

The general propensity to select relatively few links is in line with the intuition that good patterns should minimize the number of edge crossings $E_c$ \cite{purchase_metrics_2002}. The distributions of the estimated $E_c$ values actually confirm this prediction (\textbf{Fig 2c}, \textbf{Methods}).
Differently from the number of connections, the corresponding edge crossings were not statistically different between conditions (Cohen's $\lvert d \rvert \leq 0.3$). 
This result can be explained by the different velocity at which the $E_c$ values increase with the connection density. 
In random networks we demonstrated analytically and confirmed with extensive simulations that the number of edge crossings scales with the square of the connection density, i.e. $E_c = E^{max}_c \rho^2$, where $E^{max}_c$ is the maximum when the graph is completely connected (\textbf{Text S1}).
Compared to random patterns, $E_c$ increases more rapidly in long-range configurations and more slowly in short-range ones (\textbf{Fig 2c} inset).
Therefore, by opting for a different numbers of links in long-range and short-range networks, people were actually attempting to reduce the huge difference in terms of edge crossings.
Note that these results cannot be attributed to potential differences in how users navigate through the range of densities (\textbf{Fig S1}).

We next considered two representative real-world networks with nodes lying in a physical space, namely the worldwide airline network and the human connectome. For both networks, the edge weight measured the importance of the connection in terms of number of operated flights and interareal axonal fascicles (\textbf{Methods}).
By ranking the links in a weight-descending order and calculating the optimal density, the unbiased criterion automatically removed about the $95\%$ and $80\%$ of the weakest links from the airline and brain network respectively. This allowed for lighter and clearer connectivity structures as compared to the original networks (\textbf{Fig 3}). 
In addition, complementary visualizations can be obtained using Eqs 2 and ranking the links according to their actual length so to emphasize the role of short-range and long-range connectivity structures (\textbf{Fig S2-3}). 
This is particularly efficient for very large systems, unweighted interactions, and weak spatio-topological relations (\textbf{Fig S4}).

\subsection*{Generative models}

Balancing the cost of adding links and the benefit they create is at the core of spatial network modeling \cite{barthelemy_spatial_2011}.
Based on a local version of the functional $J$, we introduced a spatial growth network model that optimizes the benefit of establishing connections to hubs and the increasing cost of their length.
Specifically, when a new node $i$ arrives, a link to each of the existing nodes is created with probability

\begin{equation}
\pi_{ij}=\hat{k}_j^\alpha (1-\hat{d}_{ij})^\beta \simeq \hat{k}_j^\alpha -\hat{k}_j^\alpha \beta\hat{d}_{ij}
\end{equation}

where $\hat{k}_j$ is degree of node $j$ and $\hat{d}_{ij}$ is the distance between $i$ and $j$, both normalized by the respective maximum values in the existing network. 
Note that the higher is the degree of the target node, the higher is the cost of a link per length unit ($c=\hat{k}_j^\alpha \beta$). 
In addition, because $\hat{k}_j$ and $\hat{d}_{ij}$ are independent, both the model parameters $\alpha$ and $\beta$ are needed, thus producing a wide range of behaviors.

We first implement a simple version of the model where the average node degree remains constant. 
This is achieved by imposing that each new node has to attach a fixed number of $m$ edges, starting from an initial seed.
The $\alpha$ and $\beta$ parameters affect the degree and internode distance distributions, respectively. Short-range regular lattices are obtained for $\alpha \gg \beta$, while long-range star-like graphs are obtained when $\alpha \ll \beta$. By opportune parameter selection, Eq. 3 reduces to the uniform attachement model ($\alpha=\beta=0$) and to the preferential attachment model ($\alpha=1$, $\beta=0$) \cite{}. In these cases, the degree distributions could be analytically derived giving the typical power-law and exponential profile (\textbf{Fig 4)}. 

We next evaluate the ability of the model to reproduce the main characteristics of real networks in terms of node degree and internode distance distributions. 
We considered the brain network of the \textit{C. elegans}, for which the spatial position of the neurons, their arrival time and the synaptic connections are entirely known (\textbf{Methods}).
To reproduce the increasing average node degree during the brain development,
we implemented an hidden-variable accelerated version of Eq 3. Similar to \cite{nicosia_phase_2013}, we added the new incoming nodes according to their actual arrival time and fixed their degree $k_j$ equal to the value of the neuronal network at the adult stage.

While several trade-offs could give the real connection density, only the combination $\alpha=2.51$, $\beta=0.18$ could also accurately reproduce the node degree and link length distributions  (\textbf{Fig 5}, \textbf{Methods}). 
%
%
Notably, this goodness-of-fit could not be obtained when considering a constant cost per length unit $c=\beta$ (\textbf{Fig S5}), suggesting that the \textit{C.elegans} network has developed by pondering the cost of the connections both in terms of their length and amount.

\section*{Discussion}

Many natural, social, technological interconnected systems are characterized by a large number of connections.
By trimming edges in a way to preserve the essential properties of the original network, graph filtering is adopted in many fields from machine learning and network science, to social and biological network analysis \cite{friedman_sparse_2008,serrano_extracting_2009,tumminello_tool_2005,de_vico_fallani_topological_2017}.
From a computational perspective, graph filtering has important consequences in terms of reduced storage requirements, faster computation, and improved scalability \cite{hashemi_comprehensive_2024}. 
More in general, sparsification can be used to remove spurious or irrelevant edges, improve the accuracy of the network inference and reduce the impact of noise on the analysis \cite{peel_statistical_2022}. 

On the one hand, most graph filtering methods rely on statistical (eg, bootstrapping), topological (eg, minimum spanning tree) or combined criteria (eg, Polya filters). As such, they neglect the actual geometry of the system determined by the physical position of the nodes. On the other hand, methods that consider the actual geometry of the graph, such as the Euclidean minimum spanning tree and relative neighborhood graph \cite{toussaint_relative_1980}, generate sparse networks by keeping edges only between the spatially adjacent nodes. While the resulting skeletons match human perception, they nevertheless significantly alter the intrinsic topology, for example by constraining the actual node degree distribution.

Here, we provide a more flexible solution that seeks to preserve both the topology and geometry of the system.
This is achieved by optimally balancing the benefit of keeping the largest number of connections and the cost associated with their cumulative internode distance. The idea of maximizing the trade-off between the price for adding links and the benefit that they will create, originates from the constraints imposed by the finite resources in real-word systems \cite{barthelemy_spatial_2011,bullmore_economy_2012}. The greater the length of a connection, the more resources are required to build it. 
Here, we posit that such cost would also depend on the number of already deployed resources, i.e., the number of existing connections. This corresponds to a more careful consideration of how to utilize the limited remaining resources. In practice, the cost per length unit should not be constant but must increase with the connection density.

In terms of visualization, an increasing cost better reflects the occurrence of edge crossings, which tend to increase with longer edges and significantly boost as the number of connections grows. Specifically, we show that the number of intersections in a random graph displayed on a plane scales with the square of its connection density. As a result, the entire spectrum of edge crossings can be derived analytically by simply knowing the maximum when the graph is complete. By avoiding computationally intense heuristic calculations, this basic result may be further exploited to address open questions in graph theory and computational geometry \cite{eppstein_graph-theoretic_2010}.
To create clear and visually pleasing network visuals, it is essential to reduce long-range connections, as they often result in confusing intersections. Our filtering approach supports this principle by naturally displaying fewer (or more) links depending on whether they connect distant (or adjacent) nodes, respectively.
This result matches the central tendency of people when they are asked to add connections until the network becomes too confusing and establish unbiased criteria for achieving legible wiring patterns. Despite the existence of objective trends, the related variability among individuals indicates an intrinsic subjectivity in the human perception. This dichotomy can be found in other contexts, including modern art and aesthetics, where factors like education, culture, and personal experience can result in considerable deviation from objective criteria \cite{arnheim_art_2004}.

Our approach offers an alternative interpretation of spatial network modeling which currently only takes into account internode distances, but not the number of existing local connections\cite{barthelemy_spatial_2011}. By allowing incremental penalization costs, our general model prevents extreme "rich-get-richer" effects, which might be unrealistic or at least unfeasible in many real situations with limited resources \cite{dsouza_emergence_2007,li_limited_2022}. This aspect is not only observed in spatial networks but also in social systems where the number of possible contacts is constrained by human cognitive limits \cite{dunbar_neocortex_1992}.
We validated our hypothesis by showing that a progressive cost is essential to reproduce the main spatio-topological features of the \textit{C.elegans} neuronal network such as the presence of expensive long-distance connections \cite{kaiser_nonoptimal_2006}. 
Although our model shares similarities with other spatial network mechanisms, such as power-law economical growths \cite{barthelemy_spatial_2011,nicosia_phase_2013}, the primary goal here was to highlight the impact of incorporating a joint distance-degree penalization as compared to considering distance alone. 
Notably, the specific shape of Eq 3 establishes an equivalence between models based on attachment probabilities $\pi_{ij} \propto \hat{k}_j f(\hat{d}_{ij})$ and benefit-cost optimization $z_{ij} \simeq \hat{k}_j -g(\hat{k}_j,\hat{d}_{ij})$.

To conclude, we propose a general criterion to visualize and analyze spatial interconnected systems that leverages both topological and geometrical properties. 
We hope that our work will inspire new insights across disciplines from computational geometry and data visualization to network and cognitive sciences.

\section*{Material and Methods}

\subsection*{NetViz experiment and data analysis}

We developed NetViz as an online graphical user interface to allow people interactively visualizing and selecting an arbitrary number of connections in a network \url{https://netviz.icm-institute.org}. 
The nodes are located on a $7$ by $7$ two dimensional unitary grid and randomly shifted from their original position by a tiny factor $0.08 \times U(0,1)$. The resulting number of nodes ($N=49$) ensures an optimal visual perception from a human perspective \cite{yoghourdjian_scalability_2021}. 
To reproduce the different geometric configurations, the edges are ranked based on their length given by the internode Euclidean distance. From the longest to the shortest (long-range), from the shortest to the longest (short-range), and completely random.

Each trial starts with a preliminary window explaining the goal and conditions of the study, with no mention about the different geometric ranking. Then, a second window opens with all the nodes prompted on the screen and the software randomly selects one of the three conditions. At this point, only the first edge is displayed according to the ranking, and a cursor slider controlling the connection density is provided\footnote{The slider sensitivity allows a maximal resolution so that edges can be explored one by one.}.
Users are explicitly asked to "\textit{use the slider to keep adding connections until the graph becomes too confusing}" and confirm their choice. The count of the edges is never displayed numerically.
At the end of each trial, NetViz records the final number of retained edges as well as all the explored values; the start and end time; the user identifier and country from the IP adress; the type of configuration (\textbf{Text S2}). 

A total of $n=10687$ users from $58$ different countries have participated to the survey. Anyone capable of reading and controlling the cursor screen, with access to internet, could participate. Participants have been recruited via mailing lists, social media (Linkedin, X), printed flyers, personal contacts and via the Prolific platform specialized for gathering reliable human responses  (\url{prolific.com}) (\textbf{Text S2}).
To improve the reliability of the collected answers we filtered the data according to the following excluding rules: number of retained edges $L$ outside the range $[2,1175]$; elapsed  time outside the range $[1, 3600]$ seconds. This resulted in a dataset of $n=9610$ users. Among those, a negligible portion ($3\%$) has played the game at least two times.

The data from the NetViz experiment were used to calculate of the $\alpha$ parameter. To this end, we considered the least square error between the theoretical connection density (Eqs 2) and the mean obtained from the real users' choices $\sum_k (\rho^{theo}_k-\langle\rho^{real}_k\rangle)^2 $ where $k=\{short,rand,long\}$.
To find the optimal $\alpha^*$ that minimized the error we used a numerical interior-point method with termination tolerance $10^{-6}$ \cite{byrd_interior_1999}. Because the NetViz graphs were in average sparse, we bounded the search in the $\alpha=[0,1]$ interval and fixed to $0$ the starting point. By definition, the final filtering constant $\phi^*=\alpha^*\sqrt{N}$.

Since the number of edge crossings was not computed online, we adopted an offline reverse-engineering approach using  the actual number of links selected by the users. 
First, for each configuration, we simulated $100$ different graphs using the same NetViz layout.
Then for each connection density value (i.e., $1176$ links) we calculated the actual number of edge crossings using the Bentley-Ottman algorithm \cite{bentley_algorithms_1979}.
By averaging across samples, we then established a $1$-to-$1$ mapping between any number of links and the related edge crossings $E_c$ in each spatial configuration.

\subsection*{Real-world network data}

All the real networks used to validate our results were gathered from freely available resources.
The airline network was gathered from the OpenFlights/Airline Route Mapper Route Database (\url{https://openflights.org/data.html}). Nodes correspond to airports worldwide and links to routes between nodes. The original network is directed because of the presence of few one-way flights. For the sake of simplicity, we symmetrized the corresponding adjacency matrix (i.e., $A+A'$) and removed any isolated node. The final parsed network consisted of $N=3214$ airports and $L=18859$ weighted undirected connections. The edge weight correspond to the number of operated flights. Each node has a physical location that could be used for geographical representations in 2D or 3D.

The human connectome was obtained from the USC Multimodal Connectivity Database \url{http://
umcd.humanconnectomeproject.org}.
The database consists of $171$ connectomes obtained from healthy individuals from diffusion weighted magnetic resonance imaging (dwMRI). Structural connectivity between macro regions of interest (ROIs) has been obtained using anatomical fiber assignment through the continuous tracking (FACT) algorithm. 
The final parsed network consisted of $N=188$ brain regions and $L=5446$ weighted undirected connections. The edge weight corresponds to the group-averaged number of anatomical fibers between nodes. Each node has a physical location that corresponds to the 3D location in the standardized MNI152 brain template \cite{brown_connected_2016}.

The neuronal network was obtained from the map of the \textit{C.elegans} connectome, consisting of $279$ somatic neurons interconnected through $6393$ chemical synapses, $890$ gap junctions, and $1410$ neuromuscular junctions \cite{varshney_structural_2011}. 
Because gap junctions often overlap with synapses and synaptic connections often are reciprocated, we considered only the backbone network, in which all the synapses and gap junctions between each pair of neurons are represented by a single undirected edge. The final network obtaining a graph with $N=279$ neurons and $L=2287$ unweighted links (neuromuscular connections were excluded). Information about the growth of the neuronal network, particularly the spatial position and exact time of birth of each neuron, was reconstructed from recent literature \cite{varier_neural_2011}.

\subsection*{Model parameter selection and goodness-of-fit}
To determine the best parameter combination reproducing the topological and spatial properties of the neuronal network, we adopted a two-step procedure.
In the first step, we aimed to find which combination reproduced the actual number of connections $L$ of the \textit{C.elegans}. To do so, we considered a same broad interval for $\alpha$ and $\beta$ consisting of $1000$ values logarithmicaly spaced between $10^{-3}$ and $10^3$. Because of the intrinsic stochastic nature of the model, we simulated $30$ networks for each parameter combination and computed their average number of links $\langle L_{sim} \rangle$. We finally computed the relative error $\epsilon_{dif}=\lvert \langle L_{sim} \rangle - L \rvert/L $.

In the second step, we aimed to identify among all the possible optimal solutions, the one that best reproduced the node degree $P(k)$ and edge length distribution $P(d)$.
To do so, we considered all the parameter combinations that gave $\epsilon_{dif}<0.05$ corresponding to differences less than $5\%$ percent. Next, we adopted a particle swarm optimization bounded by the found limits. For each parameter combination we simulated $100$ networks and computed their average node degree $\langle P(k)_{sim} \rangle$ and edge length distribution $\langle P(d)_{sim} \rangle$. Finally, the evaluating function was $\epsilon_{div}=max(JS_{P(k)},JS_{P(d)})$, where $JS$ is the Jensen-Shannon divergence between the average simulated and real distributions. Other main parameters were $20$ particles, $400$ iterations max and $0.001$ tolerance.

\section*{Acknowledgments}
We would like to thank the DSI of the Paris Brain Institute for the development of the online NetViz survey; Vincenzo Nicosia for having provided insightful suggestions on the neuronal network of the \textit{C. elegans}; Marc Barthelemy, Mattieu Latapy and Lionel Tabourier for their valuable input on edge crossings.
FDVF acknowledges support from the European Research Council (ERC) under the European Unions Horizon 2020 research and innovation program (Grant Agreement No. 864729)
\nolinenumbers

\par\null

\clearpage
\bibliographystyle{plos2015}
\nolinenumbers
\begin{small}
\bibliography{netviz.bib}
\end{small}

\newpage
\begin{figure}[h!]
\begin{center}
\includegraphics[width=.66\columnwidth]{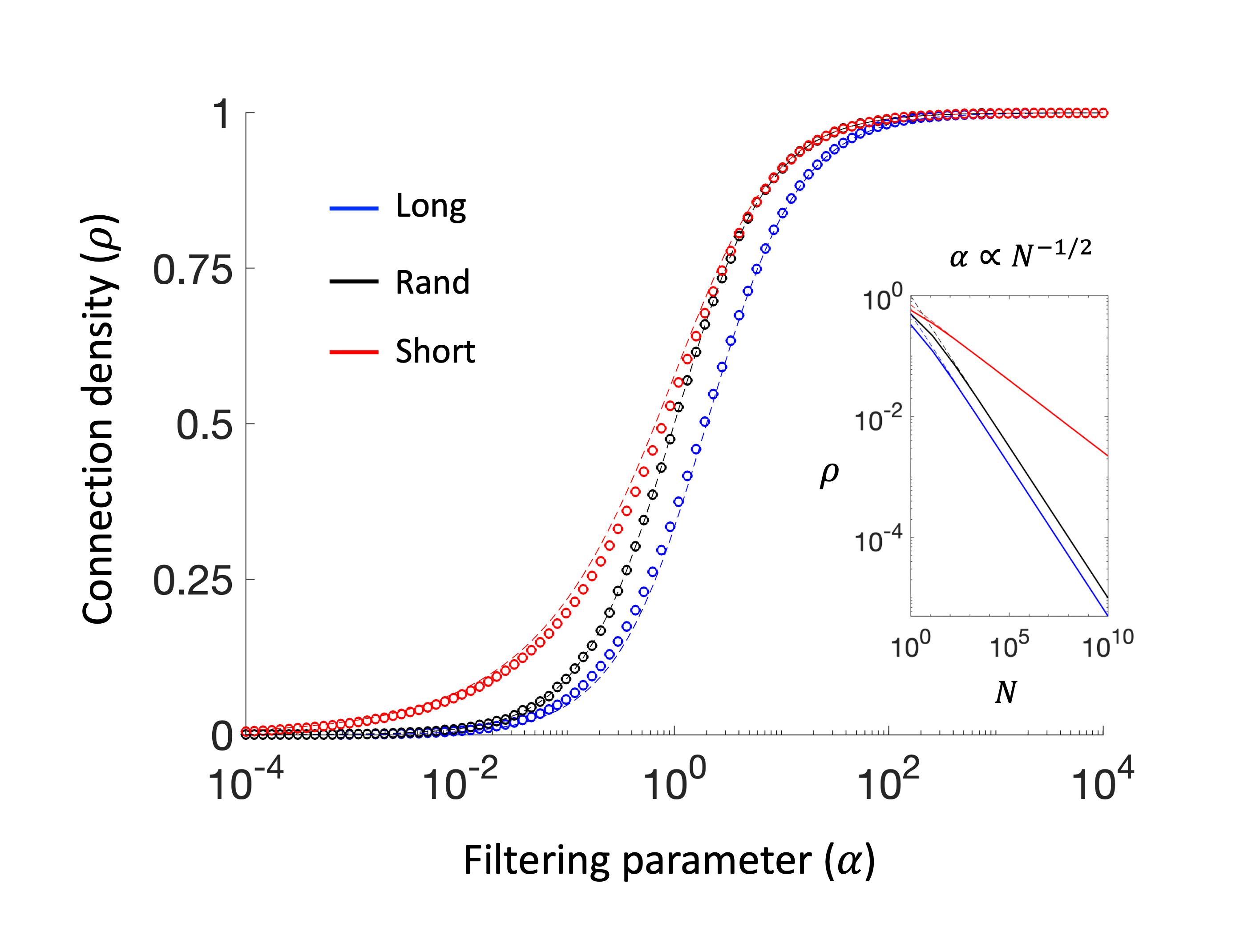}
\small{
\caption{{\bf Theoretical versus empirical behavior of the optimal connection density.}
Dashed lines represent the analytical solutions as a function of the filtering parameter $\alpha$. (Eqs 2).
Circle markers represent the values obtained by simulating a synthetic network with N=1000 nodes arranged on a pseudo-random 2D unitary circle. Black curves = no correlation between the links’ weights and lengths (rand). Blue curves = positive correaltion between the edge weights and lengths (long). Red curves = negative correaltion between the edge weights and lengths (short). 
The inset illustrates the scaling of the solution when the filtering parameter is proportional to the typical internode distance in 2D. The inset shows the case $\alpha=1/\sqrt{N}$.
Since these results only depend on the edge weight-length correlation, they stay qualitatively similar regardless of the spatial dimension, the number and position of the nodes (data not shown here).
}
}
\label{fig1}
\end{center}
\end{figure}

\newpage
\begin{figure}[h!]
\begin{center}
\includegraphics[width=1\columnwidth]{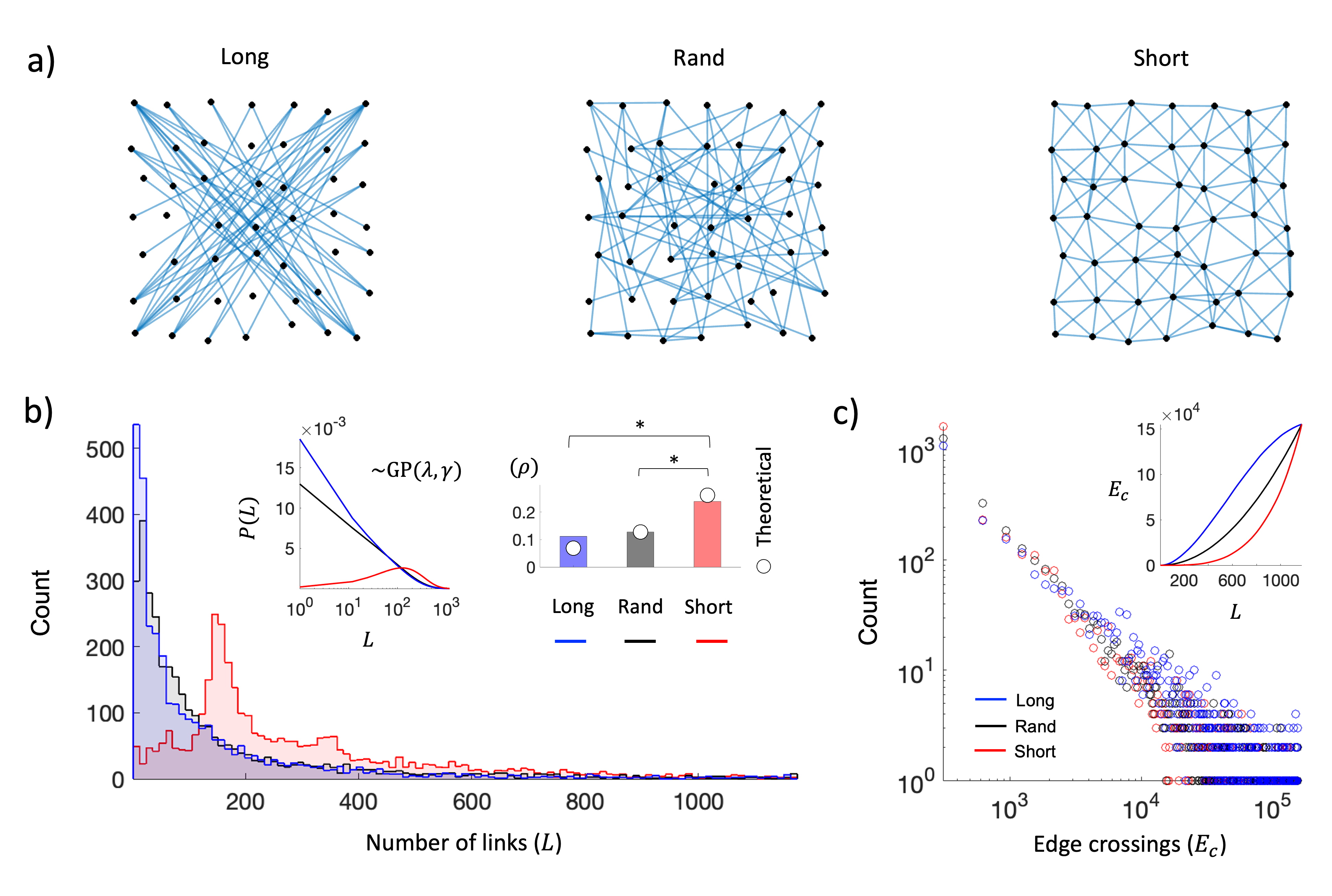}
\small{
\caption{{\bf Results from the NetViz experiment.}
a) Examples of networks displayed by NetViz. Depending on the type of configuration, the edges tend to connect the farthest nodes (long), the closest ones (short) or they are randomly distributed (rand). While visually the number of edges look similar across conditions they are instead rather different i.e., $L_{long}=43$, $L_{rand}=73$, and $L_{short}=164$. 
b) Histograms show the distributions of the number of edges selected by the NetViz participants in different conditions. The number of users in each category is $n_{short}=3322$, $n_{rand}={3154}$, $n_{long}=3163$. 
The first inset shows the count of the links as modeled by a Gamma-Poisson(GP) process with mean $\lambda$ and scale parameter $\gamma$ accounting for the overdispertion of the data (\textbf{Text S2}).
Vertical bars in the second inset show the group-averaged connection density in each condition. Asterisks indicate a significant mean-difference effects size (Cohen's $\abs{d} >0.6$). 
In the short-range condition the average number of links ($L_{short}=281.87$) is statistically higher than random ($L_{rand}=150.27$, Cohen's $d=0.6276$) and long-range ($L_{long}=131.83$, Cohen's $d=0.7314$). No statistical differences between $L_{rand}$ and $L_{long}$ (Cohen's $d=0.0927$).
White circles correspond to the theoretical conneciton densities from Eqs 2 with $\alpha=0.146$ (i.e. $\phi=1.026$) (\textbf{Methods}).
c) Distributions of the number of edge crossings $E_c$ corresponding to the number of links selected by the users in the different conditions. No significant mean-difference effects sizes between conditions (Cohen's $\abs{d}<0.3$). The inset shows the estimated $E_c$ associated with each value of $L$ in the NetViz layout in the three conditions (\textbf{Methods}).
}
}
\label{fig2}
\end{center}
\end{figure}

\newpage
\begin{figure}[h!]
\begin{center}
\includegraphics[width=1\columnwidth]{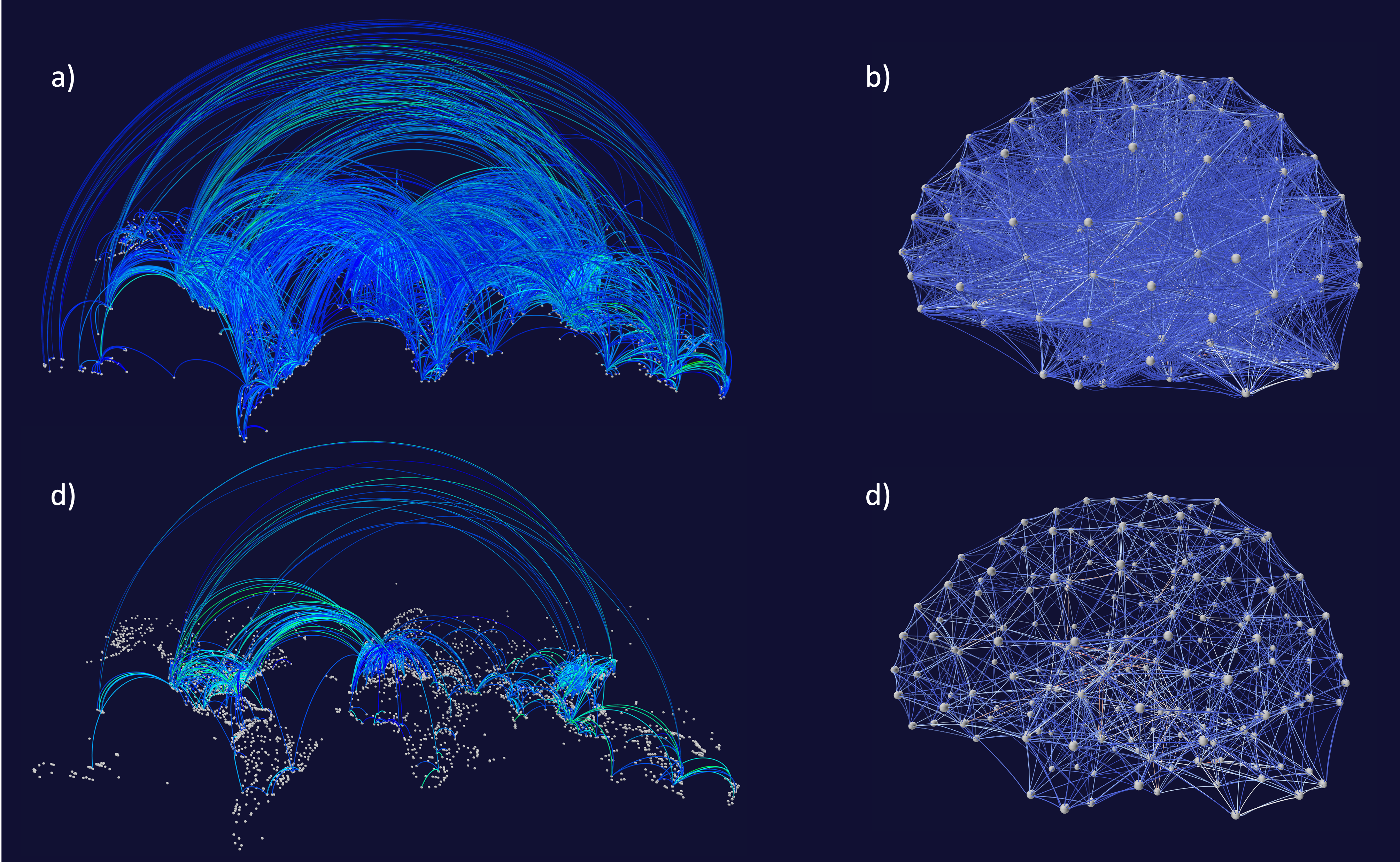}
\small{
\caption{{\bf Real-world networks filtered with the unbiased criterion.}
a) The airline route network ($N=3214$, $L=18859$). Nodes correspond to airports and links correspond to the number of operated flights. The link weight is coded by the color. The lighter the color, the higher the number of flights. For illustrative purposes, the network is shown on its 2D geographical representation. The height of the connections is proportional to the geodesic distance between the connected airports. 
b) The human connectome ($N=188$, $L=5446$). Nodes correspond to different brain regions, and links measure the number of axonal fibers between different regions (in log scale). The link weight is coded by the color. The lighter the color, the higher the number of fibers.
c) Airline route network filtered with $\phi=1$ and $s=3$ (i.e., $\alpha=1/N^{1/3}$). The optimal connection density is obtained by maximizing $J$ and sorting the links by their actual weight (descending order). Final number of filtered connections $L=915$. Around $95\%$ of the weakest connections are removed allowing to clearly visualize the main airline routes between continents.
d) The human connectome filtered with $\phi=1$ and $s=3$ (i.e., $\alpha=1/N^{1/3}$). The optimal connection density is obtained by maximizing $J$ and sorting the links by their actual weight (descending order). Final number of strongest filtered connections $L=1077$. Around $80\%$ of the weakest connections are removed allowing to clearly visualize the strong connectivity of the subcortical regions (e.g., dorsal pallidum, caudate nucleus, thalamus).
All visualizations are realized with the freely available online software VIZAJ \cite{rolland_vizaj_2022}
}
}
\label{fig3}
\end{center}
\end{figure}

\newpage
\begin{figure}[h!]
\begin{center}
\includegraphics[width=1\columnwidth]{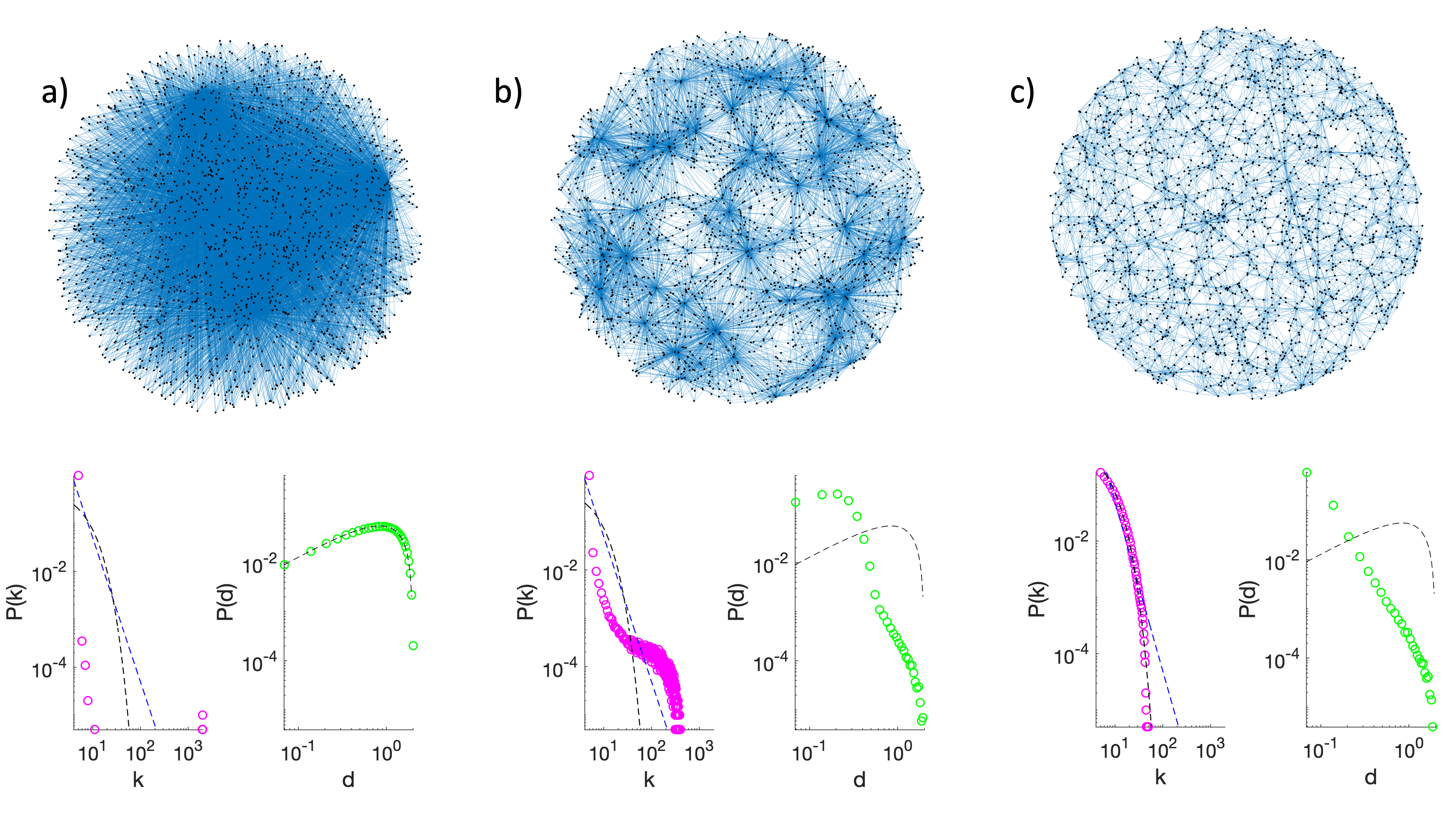}
\small{
\caption{{\bf Main features of the spatial growth network model.}
Synthetic networks are generated iteratively from an initial seed of $N_0=6$ fully connected nodes. At each step, a new node is located randomly within a unitary 2D disk and attached to $m=5$ existing nodes. The growing process stops until the total number of nodes is $N=2000$.
Three representative cases are illustrated here, according to different model parameters. The top row shows an example of the resulting network, while the bottom row reports the node degree and distance distributions averaged over $100$ samples and compared to known theoretical behaviors.
a) $\alpha=10$ and $\beta=0$. The node degree distribution (magenta circles) indicates the presence of few giant hubs and a clear difference from exponential (dashed black) or power-law (dashed blue) behavior. The distance distribution is instead perfectly matching the theoretical expectations for $N$ points randomly distributed in a unitary disk (dashed black) \cite{moltchanov_distance_2012}.
b) $\alpha=10$ and $\beta=100$. Both the node degree (dashed magenta circles) and distance (green circles) distributions show that network alters its configuration exhibiting many relatively smaller hubs and lower distances. 
c) $\alpha=0$ and $\beta=100$. The node degree distribution (magenta circles) follows a typical exponential behavior leading to more homogeneous node degrees as in Erdos-Renyi random networks (dashed blue). The distance distribution (green circles) shows that long-range connections are dramatically suppressed in favor of many short-range links. 
}
}
\label{fig4}
\end{center}
\end{figure}

\newpage
\begin{figure}[h!]
\begin{center}
\includegraphics[width=.6\columnwidth]{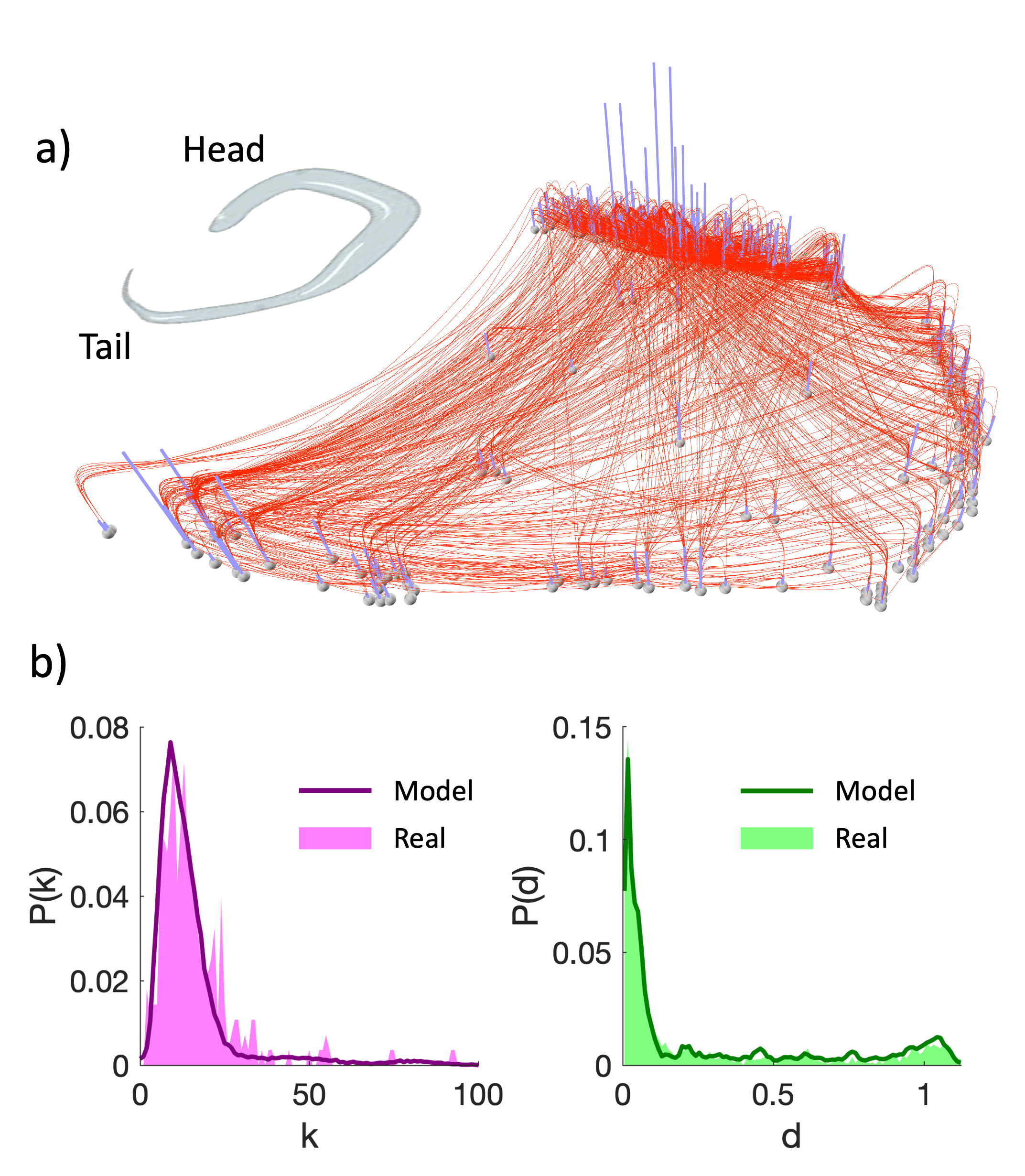}
\small{
\caption{{\bf Modeled structural and spatial properties of the C.elegans neuronal network.}
a) Representation of the C.elegans network consisting of $N=279$ neurons (grey nodes) and $L=2287$ unweighted connections (red curves). The longitudinal dimension of nodes’ location is here stretched for illustrative purposes. The height of the blue vertical bars is proportional to node degree and spot out the most connected nodes in the head of the nematode.
b) shows the node degree (magenta) and distance distributions (green) for the real neuronal network (areas) and for the ones obtained by averaging $100$ realizations of the model (solid curves). This optimal goodness-of-fit ($\epsilon_{div}= 0.118$) is obtained with the accelerated version of the spatial growth network model with parameters $\alpha=2.51$, $\beta=0.18$.
}
}
\label{fig5}
\end{center}
\end{figure}

\end{document}